\documentclass[journal]{IEEEtran}
\usepackage{amsmath,amsfonts}
\usepackage{algorithmic}
\usepackage{algorithm}
\usepackage{array}
\usepackage[caption=false,font=normalsize,labelfont=sf,textfont=sf]{subfig}
\usepackage{textcomp}
\usepackage{stfloats}
\usepackage{url}
\usepackage{verbatim}
\usepackage{graphicx}
\usepackage{cite}
\usepackage{orcidlink}
\usepackage{amssymb}
\usepackage{siunitx}
\usepackage{placeins}
\usepackage[Option]{overpic}
\usepackage{tabularx,booktabs}
\usepackage{multirow}
\usepackage{xcolor}
\usepackage[acronym]{glossaries}
\usepackage[inline]{enumitem}
\usepackage[capitalise]{cleveref}
\usepackage{svg}

\newacronym{iot}{IoT}{Internet-of-Things}
\newacronym{npu}{NPU}{Neural Processing Unit}
\newacronym{dnn}{DNN}{Deep Neural Network}
\newacronym{llm}{LLM}{Large Language Model}
\newacronym{ulp}{ULP}{Ultra-Low Power}
\newacronym{sota}{SOTA}{State-of-the-art}
\newacronym{ty}{TYv12}{TinyissimoYOLOv12}
\newacronym{fov}{FoV}{Field-of-view}
\newacronym{ai}{AI}{Artificial Intelligence}
\newacronym{mcu}{MCU}{Microcontroller Unit}
\newacronym{dvs}{DVS}{Dynamic Vision Sensor}
\newacronym{cots}{COTS}{Commercial-off-the-shelf}
\newacronym{tyv12}{TYv12}{TinyissimoYOLOv12}
\newacronym{pmic}{PMIC}{Power Managment IC}
\newacronym{sip}{SIP}{System in Package}
\newacronym{ble}{BLE}{Bluetooth Low Energy}
\newacronym{soc}{SoC}{System on Chip}
\newacronym{pcb}{PCB}{Printed Circuit Board}
\newacronym{ic}{IC}{Integrated Circuit}
\newacronym{rgb}{RGB}{Red-Green-Blue}
\newacronym{gpio}{GPIO}{General Purpose Input Output}
\newacronym{gcd}{GCD}{Global Contrast Detector}
\newacronym{ne16}{NE16}{Neural Engine}
\newacronym{hwc}{HWC}{Height-Width-Channel}
\newacronym{cnn}{CNN}{Convolutional Neural Network}
\newacronym{imu}{IMU}{Inertial Measurement Unit}
\newacronym{hmi}{HMI}{Human-Machine Interface}
\newacronym{fpga}{FPGA}{Field Programmable Gate Array}
\newacronym{ptq}{PTQ}{Post Training Quantization}
\newacronym{map}{mAP}{Mean Average Precision}
\newacronym{iou}{IoU}{Interseciton-over-Union}
\newacronym{nms}{NMS}{Non-Max-Suppression}
\newacronym{tinyML}{TinyML}{Tiny Machine Learning}

\usepackage{tikz}
\usetikzlibrary{arrows.meta}

\usepackage{eso-pic}
\usepackage{url}
\AddToShipoutPictureBG*{
  \AtPageUpperLeft{%
    \put(0,-40){\raisebox{15pt}{\makebox[\paperwidth]{\begin{minipage}{21cm}\centering
      \textcolor{gray}{This work has been submitted to the IEEE for possible publication. }
    \end{minipage}}}}%
  }
  \AtPageLowerLeft{%
    \raisebox{20pt}{\makebox[\paperwidth]{\begin{minipage}{21cm}\centering
      \textcolor{gray}{Copyright may be transferred without notice, after which this version may no longer be accessible.
      }
    \end{minipage}}}%
  }
}

\begin{document}

\title{An Energy-Proportional Multimodal and Context-Aware \\ Vision IoT Node}

\author{
Julian Moosmann\orcidlink{0009-0007-0283-0031},~\IEEEmembership{Graduate Student Member,~IEEE,}
Philipp Mayer\orcidlink{0000-0002-4554-7937},
Luca Benini\orcidlink{0000-0001-8068-3806},~\IEEEmembership{Fellow,~IEEE,}
Michele Magno\orcidlink{0000-0003-0368-8923},~\IEEEmembership{Fellow,~IEEE,}
\thanks{
		Manuscript received XXXX XX, 2025; revised XXXX XX, 2025; accepted
        XXXX XX, 2025. Date of publication XXXX XX, 2025; date of current version XXXX XX, 2025. \textit{(Corresponding author: Julian Moosmann.)}

        Julian Moosmann is with the Center for Project-Based Learning, ETH Zurich, 8092 Zurich, Switzerland, and with the Integrated Systems Laboratory, ETH Zurich, 8092 Zurich, Switzerland (e-mail: julian.moosmann@pbl.ee.ethz.ch).
		
        Philipp Mayer is with the Center for Project-Based Learning, ETH Zurich, 8092 Zurich, Switzerland (e-mail: mayerph@ethz.ch).
        
        Luca Benini is with the Integrated Systems Laboratory, ETH Zurich, 8092 Zurich, Switzerland (e-mail: lbenini@iis.ee.ethz.ch).

        Michele Magno is with the Center for Project-Based Learning, ETH Zurich, 8092 Zurich, Switzerland (e-mail: michele.magno@pbl.ee.ethz.ch).
    }
}



\maketitle

\begin{abstract}
While recent advancements in TinyML have significantly reduced the computational complexity of on-device vision pipelines, image acquisition remains a dominant contributor to system-level energy consumption and memory footprint. In vision-enabled IoT platforms, the image sensor consumes energy comparable to the inference engine, thereby offsetting algorithmic efficiency gains. Consequently, current designs face a fundamental trade-off: continuous and always-on sensing incurs prohibitive energy consumption, whereas aggressive duty cycling increases latency and risks missing transient events.

This work presents an energy-proportional, context-aware vision IoT node that addresses this challenge through a heterogeneous multimodal dual-camera architecture. Detection and recognition are decoupled by combining an event-based imager operating asynchronously in an energy-efficient always-on wake-on-motion mode together with an \acrshort{rgb} imager. Deployed on a low-power microcontroller, a novel TinyissimoYOLOv12 is introduced for efficient and accurate object detection. By activating the high-power image acquisition and processing stages only upon sparse visual triggers, the proposed architecture improves efficiency and latency, eliminating redundant sensing while maintaining continuous monitoring coverage. 
Experimental results demonstrate an energy consumption of only \SI{222}{\micro\watt h}. Upon a motion trigger, the system completes a full sense-to-report cycle---\acrshort{rgb} acquisition, object detection across 80 classes, and LoRa telemetry---with a total energy consumption of \SI{28.7}{\milli\joule}. The network achieves up to \SI{32.3}{\%} mAP with a model size of 1 million parameters. At a \SI{1}{\percent} daily activity ratio, the platform achieves a three-month operational lifetime with a \SI{1.85}{Wh} battery, enabling always-on visual monitoring in a place-and-forget scenario through autonomous edge intelligence.


\end{abstract}


\glsresetall
\section{Introduction}
\IEEEpubidadjcol
\IEEEPARstart{L}{ong-term} environmental monitoring is a fundamental requirement for emerging \gls{iot} systems, spanning infrastructure monitoring \cite{smart_cities}, smart agriculture, wildlife observation, and security applications \cite{replication_smart_city, longterm_env_monitoring, wildlife, smart_city}. Such systems must operate autonomously in a place-and-forget manner for several months or years, with limited energy availability \cite{autonomous_fire_detection}, and deliver context-aware information to the cloud \cite{ruiz2025context}. Furthermore, the ability to process and store data entirely onboard enables visual environmental monitoring in remote or rural areas with limited energy availability\cite{rural_monitoring, scharer_wind_turbine}, while preserving data privacy\cite{survey_iot}.  Achieving persistent, always-on visual context awareness\cite{ruiz2025context}---by running onboard image-processing upon sparse visual triggers---represents a critical system-level design challenge for IoT devices, as it requires the careful co-optimization of sensing, computation, communication, and power management under strict energy and resource constraints\cite{iot_frontiers, federated_learning}.
\begin{figure}
    \centering
    \includegraphics[width=0.95\columnwidth]{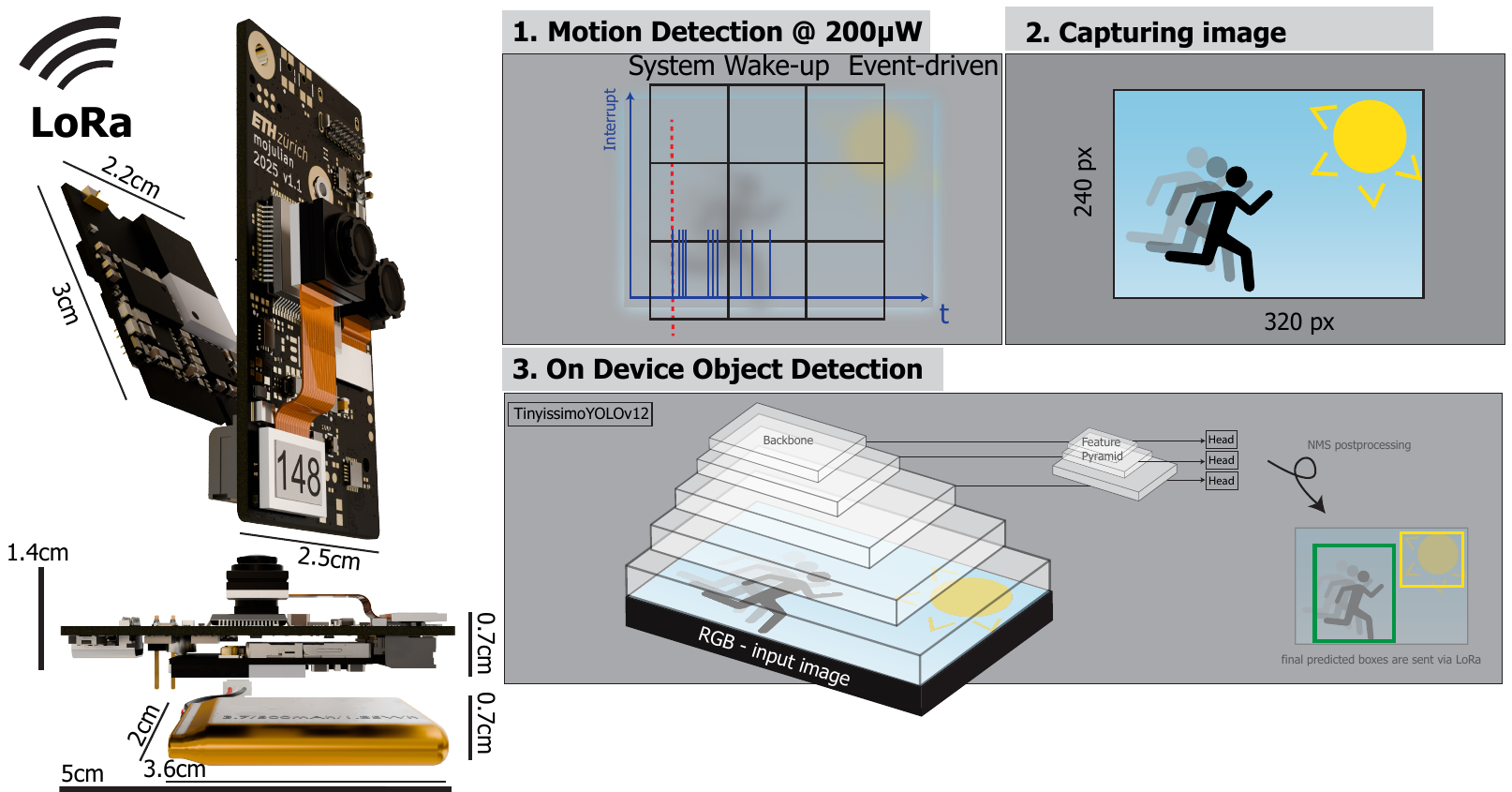}
    \caption{Prototype of the developed energy proportional and multimodal vision node with a size of \SI{50}{mm} $\times$ \SI{25}{mm} $\times$ \SI{15}{mm} }
    \label{fig:visioniotnode_rendering}
\end{figure}
Recent advancements in embedded processing, driven by novel processing units, have significantly improved energy efficiency for on-device \gls{tinyML} in a wide range of IoT applications\cite{ai_mcu, spark}. 

Typically, long-lasting and context-aware \cite{contextaware} environmental monitoring devices entail a range of sensors that consume tens to a few hundred $\mu$W---such as temperature sensors \cite{tempsensor}, \glspl{imu}, magnetometers\cite{caviezel_relevance_2021}, or microphones~\cite{9099251}---and process the sensory information using deployed \gls{tinyML} algorithms, to reduce data transmission, and prolong battery lifetime. However, camera-based IoT devices combined with \gls{tinyML} are still rarely deployed in a ``place-and-forget`` battery-operated scenario ~\cite{facedetector, 7990119, 9460037, bhattacharyya2024helios, giordano2020battery, hasan2025modeling, 11125738} primarily due to high system-level energy consumption dominated by the visual sensing modality, especially in always-on mode \cite{always_on_cam1, always_on_cam2}.  

As a result, existing visual \gls{iot} solutions relying on conventional \acrshort{rgb} cameras typically adopt one of two strategies. Either the image sensor and the associated compute units are duty cycled \cite{11125738, giordano2020battery, hasan2025modeling}, thereby reducing average power consumption at the cost of sacrificing awareness over battery lifetime; or the system operates continuously with a static power draw, limiting the battery lifetime to a few hours \cite{7990119, 9460037, bhattacharyya2024helios}. While duty-cycling lowers overall energy consumption\cite{general_duty_cycle, dutcy_cycle_example}, it introduces increased detection latency and a higher probability of missing transient events, thereby limiting its suitability for safety-critical inspection and surveillance applications \cite{always_on_cam1, giordano2020battery}.

Recent advances in neuromorphic sensing and event-based camera technology 
promise extremely low power consumption at the sensor level. Event-based cameras generate data asynchronously and sparsely, and the sensor consumes ultra-low idle power, making them well-suited for continuous visual sensing under strict energy constraints \cite{gallego2020event}. 
Despite these advantages, extracting high-level semantic information from event data remains challenging on resource-constrained devices \cite{RVT}. Most existing processing approaches for generalized object detection with event cameras target either spiking neural hardware \cite{snn_event_camera,snn_ann} or high-performance computing platforms \cite{snn_ann, leod}. When deployed on non-spiking hardware such as modern multi-core \textsc{RISC-V}-based microcontrollers equipped with \gls{cnn}-accelerators, event-based processing requires \gls{ai} algorithms that are not perfectly tailored for event data representations \cite{gallego2020event} or exceed the available computational resources \cite{RVT, chen2025eventbasedtinyobjectdetection}. Object detection and scene understanding based solely on event data remain challenging and do not achieve the same accuracy performance as \acrshort{rgb}-based \gls{sota} object detection on microcontroller-class devices\cite{bhushan2025deploying}.

Additionally, the efficiency gains of event-cameras are inherently scene-dependent: in predominantly static \glspl{fov} with sparse motion---which is common in \gls{iot} applications---, event activity remains low, and energy savings are substantial, whereas highly dynamic scenes increase event rates and diminish both power and computational advantages. This is primarily due to the sensor being energy-proportional to activity, in contrast to duty-cycled frame-based cameras, which operate energy-proportional with respect to on-time.

\begin{table*}[t]
\centering
\caption{Comparison of existing works on visual wake-up capability and system power consumption.}
\resizebox{\linewidth}{!}{%
\renewcommand{\arraystretch}{1.5}
\begin{tabular}{@{}lccccccc@{}}
\toprule
\textbf{Metric} & \textsc{IOTJ'17 \cite{7990119}} & \textsc{I2MTC'21 \cite{9460037}} & \textsc{ECCV'24 \cite{bhattacharyya2024helios}} & \textsc{ENSsys'20\cite{giordano2020battery}} & \textsc{ENSsys'25 \cite{hasan2025modeling}} & \textsc{COINS'25 \cite{11125738}} & \textbf{This Work} \\ 
\midrule
\textit{Visual Wake Up} & \SI{10}{\mu W}  & \SI{400}{\mu W} & -- & \SI{1}{mW} @ \SI{1}{Hz} & -- & -- & \SI{222}{\mu W} (ULP) \\ 
\textit{Sensor Power} & \SI{277}{\mu W} & \SI{2.4}{mW} & \SI{22.8}{mW} & \SI{40}{mW} (Active) & -- & \SI{8.1}{mW} & \SI{64.2}{\micro W} / \SI{58.8}{mW} \\ 
\textit{System Level Power} & \SI{7.62}{mW} & -- & \SI{340}{mW} & -- & -- & \SI{15.5}{mW} & \SI{63.4}{mW} \\ 
\textit{Camera Type} & DVS & Custom IC & GENX320 & BW Image & Omnivision & HM0360 & GENX320 + HM0360 \\ 
\textit{Color} & DVS & DVS & DVS & BW & RGB & RGB & \textbf{DVS + RGB} \\
\textit{Processing Platform} & FPGA + \textsc{PULPv3} \gls{soc} & FPGA: \textsc{AGLN250} &  \textsc{NXP Nano UltraLite} & \textsc{Apollo 3} vs. \textsc{SAMD51} & \textsc{MAX32660} & \textsc{GAP9} & \textsc{GAP9} \\
\textit{Data Processing} & MoG \& FD$^{i}$ & - & CNN & CNN & - & YOLOv5p & TYv12-5 \\
\textit{Algorithm Input Resolution} & 128$\times$64 & 160$\times$120 & 320$\times$320 & 30$\times$40 & - & 256$\times$256 &  256$\times$256 \\
\textit{Sample Inference Rate} & WUP + \SI{10}{FPS} & WUP + \SI{8}{FPS} & \SI{17}{FPS} & WUP + \SI{0.44}{FPS} & \SI{14.4}{min}$^{ii}$ & \SI{0.5}{FPS} & WUP + \SI{11}{FPS} \\
\textit{Network Latency} & n/a & - & \SI{60}{ms} & \SI{0.64}{s} & - & \SI{66.2}{ms} & \SI{82.7}{ms} \\
\textit{Evaluation} & Full system & Data aquisition & Full system &  Full System & Full System  & Full system & Full system \\ 
\multirow{2}{*}{\textit{Method}} & WUP + & \multirow{2}{*}{WUP} & Continuous + & \SI{1}{Hz} ToF  & Self-powered + & \SI{0.5}{Hz} &\multirow{2}{*}{WUP + TYv12-5} \\ 
 & Object Detection & & Gesture Recogn. & WUP duty-cycled & RGB LoRa trans. & duty-cycled &  \\ 
\textit{Wireless Capability} & - & - & - & LoRa & LoRaWAN / NB-IoT & BLE$^{iii}$ & BLE \& LoRa$^{iv}$\\
\bottomrule
\multicolumn{8}{l}{$^{i}$ MoG: Mixture of Gaussian; FD: Frame Difference --- both classical computer vision methods for object detection.} \\
\multicolumn{8}{l}{$^{ii}$ One image is sampled approximately every 14.4 minutes to achieve the battery-free operations. No object detection and inference is running on the \gls{mcu}.}\\
\multicolumn{8}{l}{$^{iii}$ BLE was not used to transfer the inference output to a visualization frontend.}\\
\multicolumn{8}{l}{$^{iv}$ Only LoRa was used to transfer the inference output to a visualization frontend.}\\
\end{tabular}%
 }
\label{tab:related_work}
\end{table*}

This paper presents and proposes an energy-proportional, multimodal, and context-aware \gls{iot} system that leverages an always-on event camera in \gls{ulp} mode, serving as a low-power attention mechanism that wakes up additional subsystems only when activity is detected within the \gls{fov} of the event camera. 
In addition, this work proposes an on-device, YOLOv12-based, quantized, and hardware-aware model---TinyissimoYOLOv12---for conventional \acrshort{rgb} imagers, which is activated only when visual motion is detected to extract additional semantic scene information with high accuracy, thereby enabling understanding of the visual environment. The on-board \gls{tinyML} pipeline runs on a low-power RISC-V processor with AI-acceleration. The whole system is presented and evaluated in this work. 

Experimental results demonstrate the high accuracy and the energy efficiency of the proposed solution. Such long-lasting multimodal sensor nodes enable deployment under a ``place-and-forget`` paradigm\cite{aksyuk_sensing} and support continuous visual sensing with high spatial and temporal fidelity. 

The key contributions are:
\begin{itemize}
    \item A multimodal visual IoT design combining onboard \gls{ai}-processing capabilities with context-aware environmental sensing and data-transmission.
    \item An energy-activity proportional sensing system that scales the energy consumption proportionally to the event rates in the environmental context from \SI{222}{\micro Wh} to \SI{28.7}{mJ} in active mode, achieving 3 months of battery runtime.
    \item An in-depth, always-on sensor evaluation of the event camera's visual wake-up capability.
    \item An optimized \gls{sota} detection network for low-power microcontrollers called \gls{ty} enhanced with attention layers, achieving a \gls{sota} detection accuracy of up to \SI{39.7}{\% mAP} by matching the imagers' resolution. 
\end{itemize}
The remaining parts of the article are organized as follows: \cref{sec:relatedWork} presents the relevant work in the field and compares the proposed work in context;
\cref{sec:visionNode} introduces the proposed system architecture for energy-proportional multimodal visual sensing.
\cref{sec:networkArchitecture} describes and evaluates the developed TinyissimoYOLOv12 deployed for image processing;
\cref{sec:results} presents the conducted experiments for evaluating the \gls{ulp} wake-up capability and the system's energy consumption, while \cref{sec:conclusion} concludes.


\section{Related Work}\label{sec:relatedWork}

Commonly used \acrshort{rgb} image sensors for IoT devices are power-hungry when operated continuously. For example, the latest generation \textsc{Himax HM0360} featuring VGA resolution, milliwatt-scale power consumption, and CSI-2 interface suitable for \gls{mcu}-class devices, consumes \SI{13.5}{mW} at a frame rate of 25~fps \cite{11125738}, which alone often exceeds the available power budget of battery-powered \gls{iot} devices.

To address these limitations, prior work has explored various system-level optimizations to reduce the energy consumption of visual \gls{iot} nodes, as listed in \cref{tab:related_work}. Early low-power vision systems, such as the work by Rusci et al. \cite{7990119}, propose an \gls{fpga}-based system combined with a binary pixel imager---which is nowadays known as \gls{dvs}---to enable event-triggered activation and visual monitoring with a higher-power processing stage. Although achieving energy efficiency as low as \SI{277}{\micro W}, the sensing resolution, the event-driven wake-up capability, and the data analysis are limited. In contrast, this work advances the concept by using higher-resolution components for both the wake-up functionality and the imaging part, and employs edge AI algorithms for on-board data evaluation.

More recent systems have investigated continuous operations of event-based cameras in battery-powered devices. Scherer et al. \cite{9460037} analyze the feasibility of always-on event sensing under strict energy constraints. While the work adopts a similar approach to our visual wake-up system, it remains limited to a proof-of-concept that performs operations without end-to-end image acquisition, onboard semantic processing, or object detection, which is in contrast to this work's proposed vision node.

Bhattacharyya et al. \cite{bhattacharyya2024helios} propose the \textsc{Helio} platform, which combines an event-based camera from \textsc{Prophesee}--- the \textsc{GENX320}---with embedded \gls{ai}-processing to perform low-power gesture recognition. The work targets \gls{hmi} applications for wearable smart eyewear devices, and as such, the \gls{ai}-processing is optimized for task-specific recognition rather than general-purpose visual perception. While the work combines the same event-camera with a deployed \gls{ai} algorithm, due to the usage of the \textsc{NXP iMX 8M Nano UltraLite} compute platform, their complete system requires power in the order of several hundreds of milliwatts. 

In summary, these prior works demonstrate that event-based cameras enable ultra-low-power, always-on monitoring and are well-suited for motion detection and wake-up mechanisms for \gls{iot} systems. However, the existing approaches either focus on task-specific applications, relying on high-power processing platforms, or lack an end-to-end pipeline for semantic visual perception on \gls{iot} devices. This work overcomes the limitations by integrating an event-based wake-up sensing with a complementary \acrshort{rgb} modality to enable energy-efficient continuous perception.

One such solution that leverages visual wake-up capabilities is the battery-free smart camera powered by energy harvesting, proposed by Giordano et al. \cite{giordano2020battery}. The system lifetime is extended and energy neutrality is reached by duty-cycling the wake-up sensing modality at \SI{1}{Hz}. This inherently reduces temporal awareness and increases the likelihood of missing short-lived visual events. Additionally, the deployed network is of a simple nature, performing face classification. In contrast, this article proposes a system to bridge the gap between energy neutrality and responsiveness to high-fidelity visual sensing by using a continuous visual wake-up modality combined with a low-power imager. Furthermore, instead of performing simple face detection, this work extends with \gls{sota} object detection on microcontroller-class devices. 

Another self-powered vision node that relies on energy neutrality is presented by Hasan et al.\cite{hasan2025modeling}. Their design also relies on infrequent image sampling to maintain functional operation without a battery. The system transmits image data wirelessly without onboard inference of \gls{ai}-algorithms, making communication energy a dominant factor. While the device is self-powered, the wireless transmission of the image is limiting the device's duty-cycling capability, underlining the need for on-board data processing and only sending data once critical information is extracted.

A multimodal sensing system approach is presented by Wiese et al. \cite{11125738} by combining visual and non-visual modalities with edge \gls{ai}. Although the approach improves overall sensing efficiency, the visual modality is again duty-cycled, and object detection is limited to a reduced-size YOLOv5p network for people detection, limiting the complexity with which the network has to operate. In contrast, this work proposes an optimized 80-class attention-based TinyissimoYOLOv12 network that achieves \gls{sota} detection performance compared to similar-sized networks.

Overall, these hybrid and multimodal sensing systems reduce the energy consumption by combining visual and non-visual modalities or by aggressively duty-cycling visual sensors. While effective for extending system lifetime, these strategies inherently reduce temporal awareness and limit responsiveness to short-lived events. This highlights the need for vision-centric, energy-proportional architectures that maintain continuous awareness while activating high-fidelity perception only when needed.

In addition to perception, wireless communication is one of the most energy-demanding components for \gls{iot} devices, particularly when transmitting high-bandwidth data such as images or video \cite{win_pottie}. To mitigate this additional overhead, recent work in \gls{tinyML} has focused on performing object detection directly on embedded devices\cite{bhushan2025deploying, 11125738, lin_mcunet_2020, lin_mcunetv2_2021, moosmann2023tinyissimoyolo, moosmann2023ultra}, thereby eliminating the need for continuous cloud communication and improving data privacy \cite{warden2019tinyml, mao2017survey}. 
Several studies have demonstrated the feasibility of deploying lightweight object detection networks on \gls{mcu}-based platforms\cite{design_tinyml, lin_mcunet_2020, lin_mcunetv2_2021, moosmann2023ultra, moosmann2023tinyissimoyolo, xu2022etinynet, dsort_boyle}. These approaches achieve reasonable performance in task-specific scenarios, where the number of object classes is limited, and the environmental conditions are constrained\cite{moosmann2024flexible, moosmann2023tinyissimoyolo}. However, compared to larger models deployed on more capable hardware\cite{aljahani2025rt}, their detection accuracy is superior. The limitations in memory and compute capacity available on \glspl{mcu} are directly proportional to the network's ability to operate on data from complex, feature-rich environments \cite{sengupta2025upscale}. 

Yet, processing the information on board can be more relevant for several applications than having increased detection capabilities.
Prior work shows that object detection can operate reliably on microcontroller-class devices for strictly defined applications \cite{dsort_boyle}. In contrast, when evaluating general-purpose detection tasks, accuracy degrades significantly \cite{moosmann2023ultra}. 
Work such as \textsc{PP-PicoDet} \cite{yu2021pp}, \textsc{YOLOX-nano} \cite{ge2021yolox} achieve a detection accuracy of up to \SI{30}{\%} $\text{mAP}^{50-95}$ on the \textsc{MS-COCO} \cite{lin2014microsoft} with sub one million parameters. In contrast, detection networks deployed on accelerated microcontrollers such as the TinyissimoYOLO network family \cite{moosmann2023tinyissimoyolo, moosmann2023ultra} achieve only poor performance on the \textsc{MS-COCO} dataset, while still having good detection performance on specialized target environments \cite{dsort_boyle}.
These limitations highlight the trade-off between energy efficiency and detection performance that remains a central challenge for vision-enabled \gls{iot} systems. 

This work addresses this gap by matching the available sensor resolution and improving the TinyissimoYOLO-family with an optimized \gls{sota} object detection network called \gls{tyv12}. This network is tailored for microcontroller-class devices, aiming to narrow the gap between ultra-small detection networks and \gls{sota} detection networks, for end-to-end on-device detection with up to 80 detection classes.

\section{Multimodal Vision Node}\label{sec:visionNode}
\cref{fig:visioniotnode_rendering} shows the developed context-aware vision node, comprising a multimodal vision sensor approach. An event-based camera is leveraged as a visual wake-up for continuous visual environmental sensing. Upon receiving an event, the system is awakened, an \acrshort{rgb} image is acquired, and the attention-based object detection network processes the input images to identify objects of interest.
The \cref{sec:hardware} details the hardware implemented to achieve multi-month lasting visual surveillance; \cref{sec:firmware} provides an overview of the implemented firmware. 
\subsection{Hardware}\label{sec:hardware}
The vision node comprises multiple \glspl{pcb} such as the computational-, and power-distribution-unit with multiple low-power sensors, as well as the camera interposer and debugging interface. Additionally, the hardware can be divided into an efficient and a performance compute domain, aligning with the energy-proportional sensing approach. The whole hardware is depicted in \cref{fig:visioniotnode}a) and the front- and back-side \glspl{pcb} are shown in \cref{fig:visioniotnode}b). The components are detailed below:

\begin{figure*}
    \centering
    \begin{center}
    \begin{minipage}{.70\textwidth}
        \centering
        \begin{overpic}[width=1\textwidth]{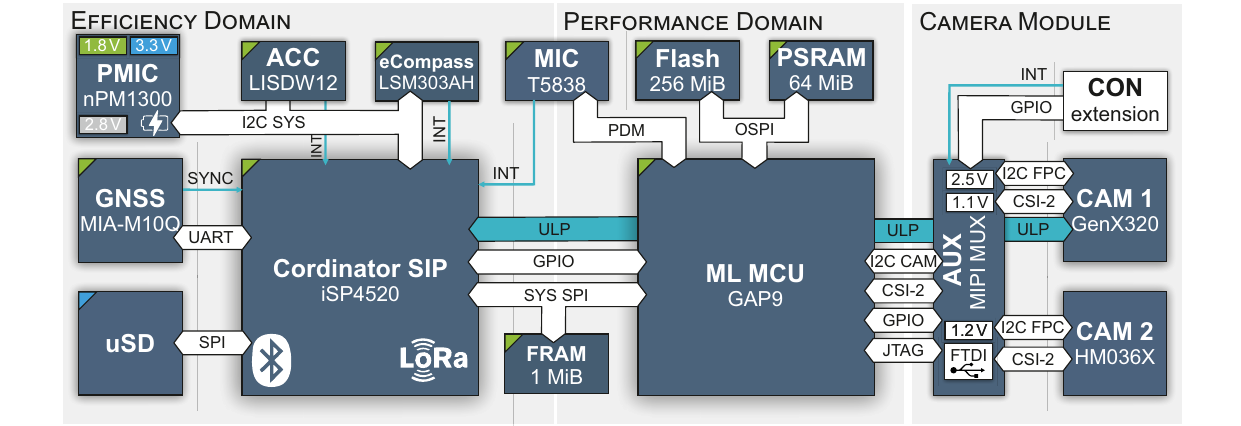}
        \put(55,-0.75){{\footnotesize(a)}}
        \end{overpic}
    \end{minipage}%
    \begin{minipage}{.30\textwidth}
        \centering
        \begin{overpic}[width=1\textwidth]{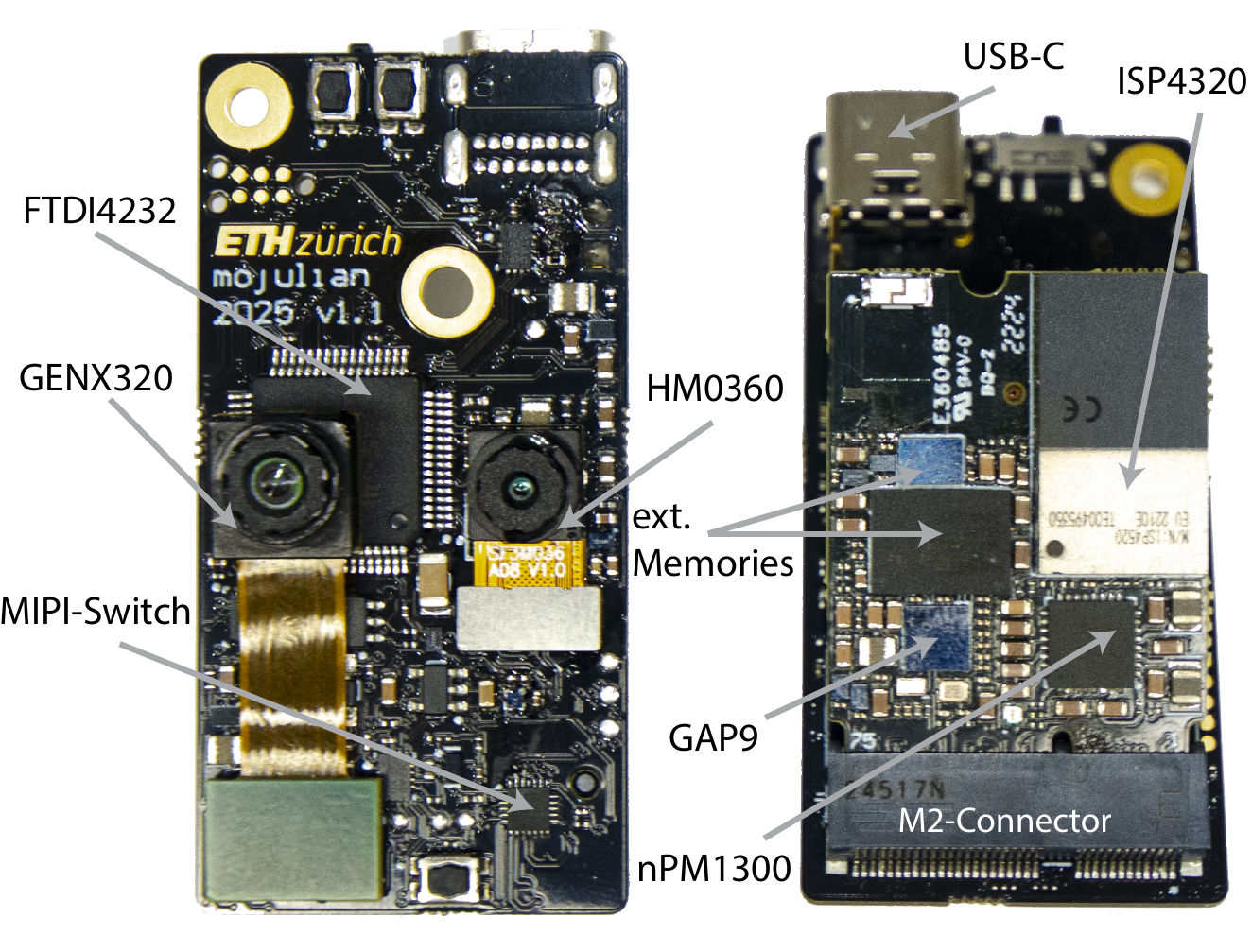}
        \put(52,-0.75){{\footnotesize(b)}}
        \end{overpic}
    \end{minipage} 
\end{center}
    \caption{Figure (a) shows the IoT node visualized as a block diagram. Color coding allows tracing of the different voltage levels. The camera module is the larger \gls{pcb} that interfaces the camera and contains debugging circuitry, while the smaller \gls{pcb} connected via an M-2 connector, contains two microcontrollers and various sensors. Figure (b) shows the annotated assembled \glspl{pcb} of the energy-proportional vision IoT node. }
    \label{fig:visioniotnode}
\end{figure*}

The compute and communication subsystem features an \textsc{nPM1300} \gls{pmic} for battery management and first-stage power distribution to the system. The \textsc{GAP9} multi-core \textsc{RISC-V} \gls{soc} is used for computational workloads, primarily for camera interfacing and \gls{ai} inference. An \textsc{iSP4520} \gls{sip} from \textsc{InsightSIP}---with built-in  \textsc{nRF52832} (\textsc{nRF52}), \textsc{LoRa} and \gls{ble}---is used to support wireless communication and ultra-low-power operation. The system runs at a nominal battery voltage of \SI{3.7}{V}. Attached to the peripherals of the \textsc{iSP4520} is a \textsc{MIA-M10Q} module from \textsc{u-Blox}, which is used for node localization, following the desired place-and-forget principle. Furthermore, an accelerometer, an e-compass, and a microphone are available and can be used for additional environmental sensing. An SD card is used for mass storage of sensory data. The two \glspl{mcu} can communicate via SPI interface while simultaneously connecting to a shared non-volatile FRAM. The shared SPI interface is used for transmitting the found boundary boxes from \textsc{GAP9} to the \textsc{nRF52}.
Additionally, the \textsc{GAP9} has access to two external memories via the processor's QSPI interface: a Flash with \SI{256}{MiB} and an additional PSRAM with \SI{64}{MiB}, which are required to run the proposed deployed TinyissimoYOLOv12 network, as the inference of the network requires a substantial amount of memory.

The camera module is the sensing subsystem, on the interposer \gls{pcb} that hosts the camera sensors and the M-2 connector, which connects to the \gls{pcb} described above. The board hosts two different camera modules. The \textsc{HM0360} from \textsc{Himax Technologies, Inc} and the \textsc{GENX320} event-camera from \textsc{Prophesee}. Both cameras are attached to the \textsc{GAP9} via the \textsc{CSI-MIPI} interface, which is multiplexed by a MIPI-switch that enables active camera selection at runtime. Additionally, each camera module includes dedicated circuitry, such as start-up sequencers, clock crystals, and \gls{gpio} extensions \glspl{ic}. Furthermore, the \textsc{GENX320} requires a level shifter for shifting the \gls{ulp} signals from \SI{2.8}{V} to \SI{1.8}{V}, enabling the ultra-low-power always-on sensing capability. 

For this work, the \textsc{GENX320} is used solely in its \gls{ulp} mode, while the \textsc{HM0360} camera is used for visually perceiving the environment. The \gls{ulp} mode of the \textsc{GENX320} groups the pixels into 9 zones of a 3 $\times$ 3 grid. Each zone aggregates the photocurrent into independent, autonomous \gls{gcd} blocks, performing relative change detection, and reporting an event by toggling the wake-up \gls{gpio} line. Additionally, an \textsc{FTDI} \gls{ic} is used for \textsc{JTAG} debugging and flashing of the \textsc{GAP9} \gls{mcu}. To fully leverage the system's low-power capabilities, the voltage rails for the two cameras and for the \textsc{FTDI} can be controlled and enabled when needed. 

\subsection{Firmware}\label{sec:firmware}
Since two \glspl{mcu} are used to run the complete system, two firmwares are developed to run in tandem. The \textsc{nRF52} is in control of the always-on low power efficiency domain---see \cref{fig:visioniotnode}, as it can operate in an \gls{ulp} mode consuming only \SI{0.7}{\micro A} during system-off with full RAM retention. On the contrary, the \textsc{GAP9}, when triggered, runs the object detection model---accelerated by its \gls{ne16} and multi-core processors---on captured RGB images. In the following paragraphs, the two firmwares for the \textsc{nRF52} and the \textsc{GAP9} are detailed:  

\textit{Efficiency Domain}: To achieve low energy consumption, the firmware of the \textsc{nRF52} is kept very basic. Once the system and all its peripherals are initialized, the system is turned off with full RAM retention, and it waits for the external \gls{gpio}-wakeup interrupt from the \textsc{GENX320} event-camera, which is in \gls{ulp}-mode. Once motion is detected by the event-camera, an interrupt is received by the \gls{mcu}, which powers and turns on the \textsc{GAP9} \gls{mcu}, and then waits for a SPI transaction to be received from the \textsc{GAP9}. Once the information is received, it is wirelessly transmitted to a relay station using \textsc{LoRa}, operating at \SI{865.1}{MHz} with a spreading factor of 7 to minimize latency and maximize energy efficiency.

\textit{Performance Domain}: The firmware of the \textsc{GAP9} adopts a sequential "race-to-halt" approach to perform inference and turn off as quickly as possible. Once the \gls{mcu} is powered on by the \textsc{nRF52}, the system and its peripherals are initialized. An image is acquired from the \textsc{HM0360} sensor, and the raw sensor data is demosaiced and white-balanced on the \textsc{GAP9} cluster. De-Bayering and white balancing are crucial for accurate object detection, as the input image must lie within the image space in which the network was trained. During demosaicing, the image's memory layout is translated into 3-channel \textsc{hwc} \acrshort{rgb} format, which is expected by the network. Thereafter, the image is analyzed by running an inference on a multi-core microprocessor accelerated by a neural engine. The network's output is post-processed by applying \gls{nms} to obtain predicted bounding boxes with assigned object classes. The filtered information is then sent via SPI to the \textsc{nRF52} \gls{mcu} for wireless communication. If no object is detected, the \textsc{GAP9} is turned off; if additional objects are detected, the firmware acquires another image of the environment for analysis. 

\textit{\textsc{GAP9} Memory allocation:} As the available L2 memory is very limited and L3 memory is energy-costly to use, the used memory layout for the buffer allocations in L2 is presented in this section. The raw captured Bayern-pattern image has dimensions 320x240 pixels, with each pixel having a color depth of \SI{8}{bits}. Demosaicing and white-balancing the raw image, a buffer of \SI{224}{KiB}\footnote{A kibibyte (KiB) is defined as 1024 bytes, while a mebibyte has 1,048,576 bytes. In our case $\frac{230'400}{1'024\text{ bytes/KiB}}= 224\text{ KiB}$} is required. Since the network expects a square image---that is, divisible by 32---a resolution of 256x256 was chosen, resulting in a buffer of \SI{192}{KiB}. At the time the network is initialized, it requires a static memory buffer allocation of \SI{118.7}{KiB} and a dynamic buffer of \SI{467.25}{KiB}. The dynamic buffer is required and functions as a scratchpad memory for the network's inference. Therefore, this buffer can be repurposed for preparing the input data. The network output is \SI{441}{KiB} in size. The size is explained by the large number of distinct detection classes (80) combined with the floating-point precision set for the network's output, $1344$x$84$x$4$. Postprocessing the network's output requires three additional buffers of size \SI{31.5}{KiB}, \SI{1344}{B}, and \SI{960}{B}. While the firmware is running sequentially, not all buffers need to remain alive for the firmware to continue. Therefore, initializing the memory at the start and reusing the available dynamic memory buffer is crucial for continuous firmware operation. For image capture and preprocessing, the available dynamic network buffer is repurposed and used in a ping-pong fashion, whereas the output buffers must be kept statically in L2.


\section{Dataset and Network Architecture}\label{sec:networkArchitecture}
This section first presents the acquired dataset used for finetuning. In addition, the developed TinyissimoYOLOv12 is compared to \gls{sota} detection networks. Lastly, the different network versions are deployed on the \textsc{GAP9} and evaluated for latency and computational efficiency. 

\subsection{Dataset Aquisition}\label{sec:dataset_acquisition}
\begin{figure}
    \centering
    \begin{overpic}[width=0.9\columnwidth]{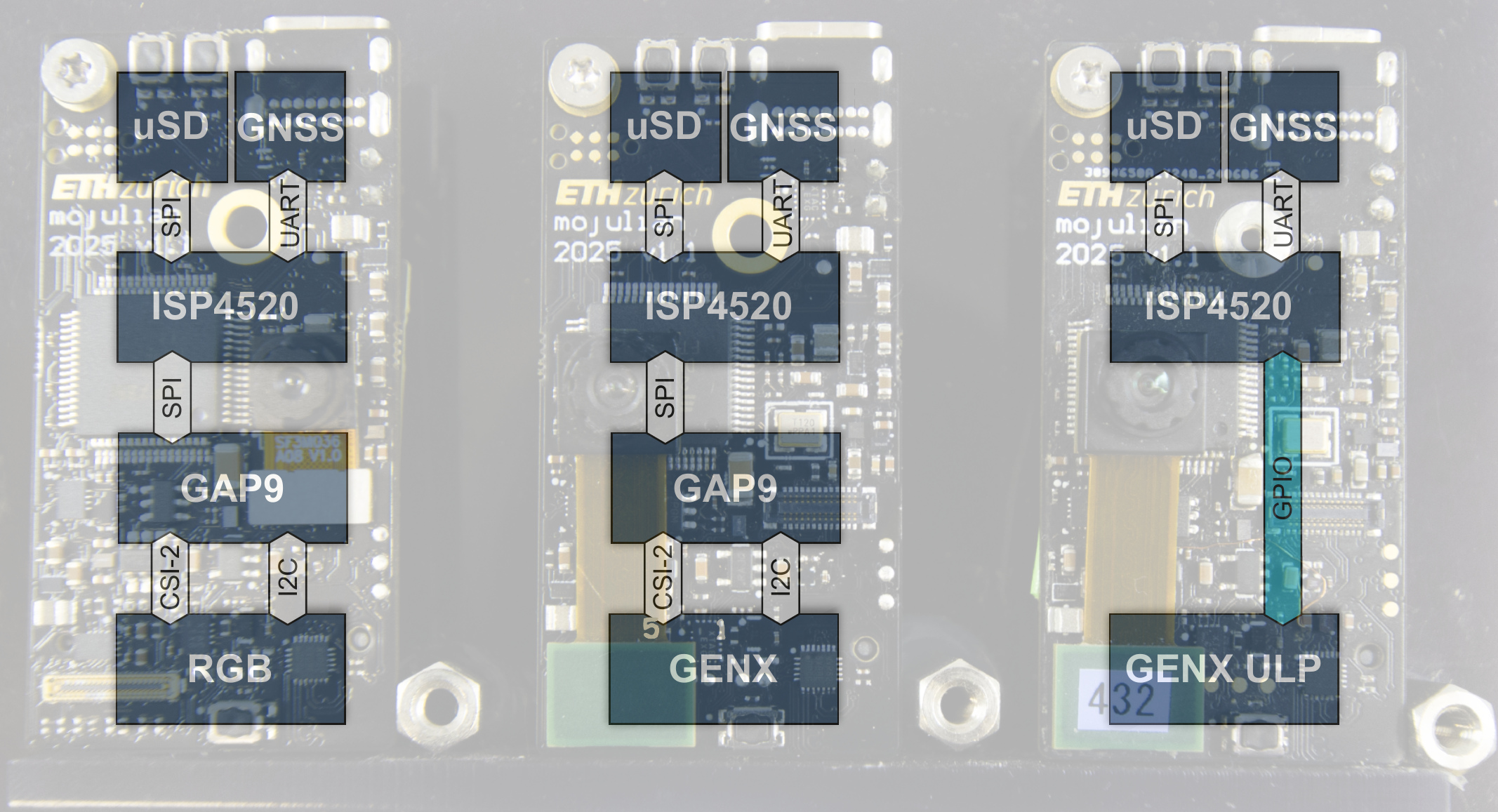}
        \put(7.5,-5){\colorbox{white}{\textcolor{black}{{\footnotesize RGB}}}}
        \put(42,-5){\colorbox{white}{\textcolor{black}{{\footnotesize GENX}}}}
        \put(78,-5){\colorbox{white}{\textcolor{black}{{\footnotesize ULP}}}}
    \end{overpic}
    \caption{Aligned and time-synchronized hardware used for data collection.}
    \label{fig:dataset_collection_hw}
\end{figure}

Training a network on a public and open-source dataset, and deploying it on custom hardware using a low-power, and low-resolution camera---which camera wasn't part of the public dataset---inherently leads to a decrease in detection robustness, as illustrated by Carlson et al. \cite{sensor_transfer}. To compensate for the camera intrinsics of the used sensor, a dataset is acquired to fine-tune the network. The data acquisition system is presented below: 

To acquire the dataset, multiple vision IoT nodes were used; see \cref{fig:dataset_collection_hw}. One node was used to acquire the \acrshort{rgb} image, the second to acquire accumulated event histograms, and the last node to collect \gls{ulp} wake-up interrupts for evaluation. The three devices were time-synchronized using the onboard GPS module. Each time a second is completed, a time pulse is sent from the GPS module to the \textsc{nRF52}, interrupting the running firmware and timestamping the data in the case of the \gls{ulp}-node, while requesting the next images from the \textsc{GAP9} in the two camera cases. Once \textsc{GAP9} acquires an image, it is sent via SPI to the \textsc{nRF52} \gls{mcu}, which in turn stores the image along with the timestamp metadata on the SD card. This enables continuous data collection outdoors and allows post-synchronization of the acquired data if needed. Additionally, the event camera and the frame-based camera have similar \glspl{fov} and matching camera extrinsics. The wide \gls{fov} of the \textsc{GENX320} camera was decisive in the custom manufacture of a \textsc{HM0360} module with a matching 130° \gls{fov} lens. During dataset collection, the light intensity was measured with a Lux Meter, the environment and scenario were documented, and the distances to the objects were recorded.

For labeling the ground-truths to the \acrshort{rgb} images and event-histograms, \textsc{Label Studio} was used. The \acrshort{rgb} images were semi-automatically labeled using a pre-trained large YOLOv8\cite{YOLOv8} network. In addition, each label was manually inspected for correctness and corrected if necessary. The event histograms were manually labeled. The dataset comprises 15'309 \acrshort{rgb} images, 26'391 event-histogram images, and time-synchronized data points for the \gls{ulp} evaluation. The detailed number of objects per sensing modality is listed in \cref{tab:detailed_dataset}.
\begin{table}[t]
\centering
\caption{Number of objects contained in the collected dataset.}
\resizebox{\linewidth}{!}{%
\begin{tabular}{@{}llllccccc@{}}
\toprule

& \multirow{2}{*}{\textbf{Person}} & \multirow{2}{*}{\textbf{Drone}} & \multirow{2}{*}{\textbf{Car}} & \multicolumn{2}{c}{\textbf{two-wheelers}} & \multicolumn{2}{c}{\textbf{big-vehicles}} &  \\ 
& & & & \textbf{Bicycles} & \textbf{Motorcycle} & \textbf{Truck} & \textbf{Bus} \\ 
\midrule
\multirow{2}{*}{\gls{dvs}}& \multirow{2}{*}{6094} & \multirow{2}{*}{2277} & \multirow{2}{*}{6106} & \multicolumn{2}{c}{1981} & \multicolumn{2}{c}{974} \\ 
& & & & 1880 & 101 & 570 & 404 \\
\multirow{2}{*}{\acrshort{rgb}} & \multirow{2}{*}{4486} & \multirow{2}{*}{5071} & \multirow{2}{*}{4586} & \multicolumn{2}{c}{391} & \multicolumn{2}{c}{659} \\ 
& & & & 353 & 38 & 420 & 239 \\
\bottomrule
\end{tabular}
}
\label{tab:detailed_dataset}
\end{table}

\subsection{TinyissimoYOLOv12}\label{sec:detection_network}
The detection network was first optimized for detection accuracy on the \textsc{MS-COCO}\cite{lin2014microsoft} dataset, then quantized and deployed on the \textsc{GAP9} \gls{mcu}. 
\paragraph{Network Optimization and Training}
To develop an optimized detection network, YOLOv12-nano was used as a starting point, as it achieves \gls{sota} detection performance while outperforming other networks with a comparably small parameter count. \cref{tab:net_optimization_operation} lists network block optimization considered for decreasing the parameter count of YOLOv12, while keeping the network's detection capability as high as possible. The first optimization replaces the bottleneck block of YOLOv12, used for blocks \textsc{C3k2} and \textsc{A2C2f}, with a SqueezeNext block followed by a depthwise convolution. The resulting network drops slightly to \SI{39.7}{\%} $\text{mAP}^{@50-95}$. The next quarter of parameters is removed by replacing the 3x3 convolutions with 3x3 depthwise-separable convolutions, followed by a 1x1 pointwise convolution. This change is inspired by the \textsc{MobileNetV1}\cite{howard2017mobilenets} architecture and results in increased network depth, with more layers. Lastly, \textsc{C3k2} and \textsc{A2C2f} are set to 1 repetition, thereby reducing the number of single blocks. Subsequent optimization steps from TYv12\_5 to TYv12\_11 are related to the detection head only. More specifically, versions 5 to 7 reduce the hidden channel depth of the A2C2f layer by a factor of 2, resulting in depths of 64, 32, and 16. Version 8 is the same as version 7, but uses 3x3 convolutions instead of depthwise-pointwise layers. Version 9 is also based on version 7 but omits the P4 part of the detection head, extracting only large and small objects. Version 10 reduces the hidden channel count in the first A2C2f layer of the network from 256 to 128, while version 11 again uses 3x3 convolutions instead of depthwise-pointwise layers. The networks are trained within the \textsc{Ultralytics}\footnote{\url{https://github.com/ultralytics/ultralytics}} framework using input resolution of 640x640. Pretraining is performed for 1000 epochs on \textsc{MS-COCO}, and finetuned for another 100 epochs on the custom-acquired dataset using a batch size of 256. Once the network is sufficiently trained, it is exported as a .onnx file with an input resolution of 256x256 and fine-tuned on the \acrshort{rgb} dataset---see \cref{tab:detailed_dataset}---for deployment.
\paragraph{Network Deployment}
To deploy the network on the target platform of \textsc{GAP9}, the network's floating-point representation is loaded into the \textsc{NNTool} to apply the available \gls{ptq} flow. The \gls{ptq} applies the statistics of the target custom dataset to the network's weights and quantizes them to 8 bits. Once the network is quantized, the \textsc{Autotiler} automatically generates C code for the network's kernels. In addition, the tool accounts for available and specified L1, L2, and L3 memory and optimizes the C code for optimal performance on the cluster core and on the accelerator \gls{ne16}. With respect to the available memory described in \ref{sec:firmware}, the available L2 buffer is set to \SI{600}{KB} for the autotiler, resulting in a static memory allocation of \SI{119}{KiB} and a dynamic allocation of \SI{468}{KiB}. The minimum size of the dynamic buffer is \SI{417}{KiB}.
\paragraph{Network Results}\label{sec:ty_results}
To compare the network's detection capability with \gls{sota} networks, the evaluated networks are trained and evaluated with an input resolution of 640x640 pixels. While this input resolution is infeasible on the current platform---since the entire input must be available in L2 memory along with the constructed network---the deployed networks use an input resolution of 256x256. While the detection capability decreases with lower input resolution, one can reduce the number of detection classes to compensate for the per-class detection capability. Yet, to showcase the feasibility of deploying an 80-class detection network, the detection head is kept for 80 classes, making memory allocation more constrained than for a network trained on fewer classes. In addition, the energy measurements are also performed with the 80-class detection head, providing an upper-bound worst-case energy consumption for the proposed system.

Once the network is trained on the \textsc{MS-COCO} dataset, it is evaluated on the corresponding validation dataset. \cref{tab:ablation_study} visualizes the achieved detection accuracy for the evaluated networks developed in the network ablation study \cref{sec:detection_network}. The smallest network with a total number of \SI{450}{k} achieves a \gls{map} of \SI{23.6}{\%} $\text{mAP}^{@50-95}$, the middle sized network with \SI{1.4}{M} parameters achieves \SI{35.1}{\%} $\text{mAP}^{@50-95}$, while the largest network with \SI{2.3}{M} parameters achieves \SI{39.6}{\%} $\text{mAP}^{@50-95}$. 

For deploying the network on the \textsc{GAP9} architecture running at \SI{370}{MHz}, the available L2 memory is set to \SI{600}{KB} to ensure sufficient memory for postprocessing the network's output. All trained and optimized networks were evaluated for inference latency and operation per cycle, as shown in \cref{tab:ablation_study} using input resolutions of 256x256. While the fastest TYv12\_11 achieves a latency as low as \SI{61.3}{ms} and a computational efficiency of up to \SI{10.1}{ops/cycle}, it is not the smallest developed network. The smallest TYv12\_10 network (\SI{450}{K} parameters) is most affected by the softmax operation in the attention layers, as the ratio between softmax and convolutional operations is highest for this model. The softmax layers must run on the cluster cores in floating-point precision, which significantly reduces computational efficiency. Lastly, the network with the highest computational efficiency is YOLOv12-small, which fits within the available memory and runs in \SI{231.4}{ms} on \textsc{GAP9}.
\begin{table}[t]
\centering
\caption{Comparison of different block types in terms of parameters per block.}
\renewcommand{\arraystretch}{1.2}
\resizebox{\linewidth}{!}{%
\begin{tabular}{@{}lccc@{}}
\toprule
\multirow{2}{*}{\textbf{Block Type}} & \multirow{2}{*}{\textbf{Parameters}} & \multirow{2}{*}{\textbf{Residual}} & \textbf{Layer breakdown} \\ 
& & & \textit{(input $\rightarrow$ hidden $\rightarrow$ output channel)} \\ 
\midrule
\multirow{2}{*}{Conv2D 3$\times$3} & \multirow{2}{*}{36,864} & \multirow{2}{*}{\textcolor{red}{$\times$}} & Standard convolution \\ 
 & & & \textit{(64 $\rightarrow$ 64)} \\
\multirow{2}{*}{SqueezeNext\cite{squeezenext}} & \multirow{2}{*}{13,312}& \multirow{2}{*}{\textcolor{green}{$\checkmark$}} &
1$\times$1 $\rightarrow$ 3$\times$3 $\rightarrow$ 1$\times$1 \\ 
& & & \textit{(64 $\rightarrow$ 32 $\rightarrow$ 64)} \\
\multirow{2}{*}{MobileNetV1\cite{howard2017mobilenets}} & \multirow{2}{*}{4,672} & \multirow{2}{*}{\textcolor{red}{$\times$}} & 3$\times$3 DW $\rightarrow$ 1$\times$1 PW \\ 
& & & \textit{(64 $\rightarrow$ 64)} \\
\multirow{2}{*}{MobileNetV2\cite{mobilenetv2}} & \multirow{2}{*}{52,608} & \multirow{2}{*}{\textcolor{green}{$\checkmark$}} & 1$\times$1 Expand $\rightarrow$ 3$\times$3 DW $\rightarrow$ 1$\times$1 projection \\ 
& & & \textit{(64 $\rightarrow$ 384 $\rightarrow$ 64)} \\
\multirow{2}{*}{MobileNetV4\cite{mobilenetv4}-style} & \multirow{2}{*}{163,840} & \multirow{2}{*}{\textcolor{red}{$\times$}} & 1$\times$1 Expand $\rightarrow$ 3$\times$3 Conv \\ 
& & & \textit{(64 $\rightarrow$ 256 $\rightarrow$ 64)} \\
\multirow{2}{*}{GhostNet\cite{ghostnet}} & \multirow{2}{*}{30,048} & \multirow{2}{*}{\textcolor{green}{$\checkmark$}} & Ghost Bottleneck ($t=6$, $s=2$) \\ 
& & & \textit{ (64 $\rightarrow$ 384$\rightarrow$ 64)} \\
\bottomrule
\end{tabular}
}
\label{tab:net_optimization_operation}
\end{table}

\begin{table*}[t]
\centering
\caption{Ablation study on YOLOv12 nano modifications for developing suitible TinyissimoYOLOv12 variants.}
\renewcommand{\arraystretch}{1.3}
\resizebox{\linewidth}{!}{%
\begin{tabular}{@{}lllllllrc@{}}
\toprule
\multirow{2}{*}{\textbf{Acronym}} & \multirow{2}{*}{\textbf{Modification}} & \multirow{2}{*}{\textbf{Parameters}} & \multirow{2}{*}{\textbf{Layers}} & \multicolumn{3}{c}{\textbf{mAP@50:95}} & \multirow{2}{*}{\textbf{latency} (ms)} & \textbf{Computational Efficiency} \\ 
& & & & 640x640 & 416x416 & 256x256 & & (ops/cycl.) \\
\midrule
 & \textit{Baseline: }YOLOv12 small \cite{tian2025yolov12} & \SI{9.2}{M} & 272 & \SI{48}{\%} &  & & 231.4 & 20.4\\ 
 & \textit{Baseline: }YOLOv12 nano \cite{tian2025yolov12} & \SI{2.6}{M} & 272 & \SI{40.6}{\%} &  & & 101.4 & 14.4\\ 
\multirow{2}{*}{TYv12\_1-2.3M} & Bottleneck Block inside C3k2, A2C2f & \multirow{2}{*}{\SI{2.38}{M}} & \multirow{2}{*}{200} & \multirow{2}{*}{\SI{39.7}{\%}} &  & & \multirow{2}{*}{108.7} & \multirow{2}{*}{12.2}\\
& SqueezeNextBlock+DW Conv & & & \\ 
TYv12\_2-1.8M & Change Conv; if c1==c2 shortcut & \SI{1.80}{M} & 365 & \SI{37.8}{\%} &  & & 106.8 & 9.5\\ 
\multirow{2}{*}{TYv12\_3-2.0M} & C3k2 to half Repeats and A2C2f to half Repeats + 1 & \multirow{2}{*}{\SI{2.0}{M}} & \multirow{2}{*}{297} & \multirow{2}{*}{\SI{37.1}{\%}}  &  & & \multirow{2}{*}{93.1} & \multirow{2}{*}{12.9}\\ 
& 3$\times$3 DW $\rightarrow$ 1$\times$1 PW & & & & & & & \\
TYv12\_4-1.4M & A2C2f to half Repeats (all layers once, dw\_pw conf) & \textbf{\SI{1.4}{M} $\downarrow$} & \textbf{309 $\uparrow$} & \textbf{\SI{35.1}{\%} $\downarrow$} & 29.5 & 26.0 & 88.4 & 10.0\\
TYv12\_5-1.0M & 2nd and 3rd A2C2f hidden layer divided by 2 & \textbf{\SI{1.0}{M} $\downarrow$} & \textbf{309} & \textbf{\SI{32.3}{\%} $\downarrow$} & 28.0 & 24.9 & 82.7 & 9.2\\
TYv12\_6-0.985M & 2nd and 3rd A2C2f hidden layer divided by 4 & \textbf{\SI{985}{K} $\downarrow$} & \textbf{309} & \textbf{\SI{32}{\%} $\downarrow$} & & & 99.6 & 6.7 \\
TYv12\_7-0.834M & 2nd and 3rd A2C2f hidden layer divided by 8 & \textbf{\SI{834}{K} $\downarrow$} & \textbf{309} & \textbf{\SI{29}{\%} $\downarrow$}  & & & 95.5 & 6.1\\
TYv12\_8-1.2M & 2nd and 3rd A2C2f hidden layer divided by 8 + 3x3 conv & \textbf{\SI{1.2}{M} $\uparrow$} & \textbf{216 $\downarrow$} & \textbf{\SI{32.4}{\%} $\downarrow$} & & & 88.3 & 10.2\\
TYv12\_9-0.722M & again 3$\times$3 DW $\rightarrow$ 1$\times$1 PW  + less residual in head & \textbf{\SI{722}{K} $\downarrow$} & \textbf{220 $\uparrow$} & \textbf{\SI{27.5}{\%} $\downarrow$} & 22.2 & 19.3 & 75.9 & 5.0\\
TYv12\_10-0.450M & 1st A2C2f hidden layer divided by 2 & \textbf{\SI{450}{K} $\downarrow$} & \textbf{220} & \textbf{\SI{23.6}{\%} $\downarrow$} & 19.3 & 16.9 & 66.4 &  4.9 \\
TYv12\_11-0.767M & 3x3 Conv & \textbf{\SI{767}{K} $\uparrow$} & \textbf{220} & \textbf{\SI{27.4}{\%} $\downarrow$} & 23.5 & 20.6 & 61.3 & 10.1\\
\bottomrule
\end{tabular}
}
\label{tab:ablation_study}
\end{table*}

\section{Results and Discussion}\label{sec:results}
This section presents the conducted experiments and the achieved energy results. First, the always-on wake-up capabilities of the used \textsc{GENX320} camera in \gls{ulp} mode is presented in \cref{sec:ulp_results}. Second, the whole energy consumption of the system experiments is analysed in \cref{sec:energy_results}.

The tight integration of the entire electrical subsystem prevents measuring the current consumption of each voltage rail individually. Therefore, the \textsc{power profiler kit II } from \textsc{Nordic Semiconductor} is used to provide a nominal battery voltage and to measure the system current directly at the battery supply. The dedicated \gls{gpio} pins of the power profiler are used to synchronize the \gls{ulp} wakeups from the \textsc{GENX320} camera directly with the power measurements. Additionally, to measure the current of the \SI{2.5}{V} rail that powers the \textsc{GENX320} camera in its \gls{ulp} mode, a \SI{0}{Ohm} jumper resistor is removed, and the same power profiler is used to perform in situ current measurement. No notable current increase or decrease occurs, with many or no events happening, which is why only the average current consumption is reported. 

\subsection{ULP Detection Capabilities}\label{sec:ulp_results}
To evaluate the performance of the visual wake-up functionality in the \gls{ulp}-mode of the \textsc{GENX320}, an experiment is conducted to measure the range at which objects trigger the event-camera's wake-up mode. These experiments were conducted indoors and outdoors at variable distances. The light conditions ranged from \SI{200}{lux} indoors to \SI{20}{klux} under cloudy conditions and \SI{60}{klux} under sunny conditions.
Furthermore, the chosen lens with which the \textsc{GENX320} is pre-equipped has a diagonal \gls{fov} of 104°. Two different object classes---person and person with bicycle---were used for the evaluation, in indoor and outdoor scenarios at different lighting conditions ranging from \SI{200}{lux} indoors to \SI{20}{klux} outdoors in cloudy weather and \SI{60}{klux} in sunny weather. To test the \textsc{GENX320} wake-up capability, the objects are purposely moved across the \gls{fov} to check whether or not a wake-up is triggered. Each object's distance to the lens is tested 10 times. 
\begin{figure}
    \centering
    \includegraphics[width=0.95\columnwidth, trim=2.35cm 1.25cm 0 0, clip]{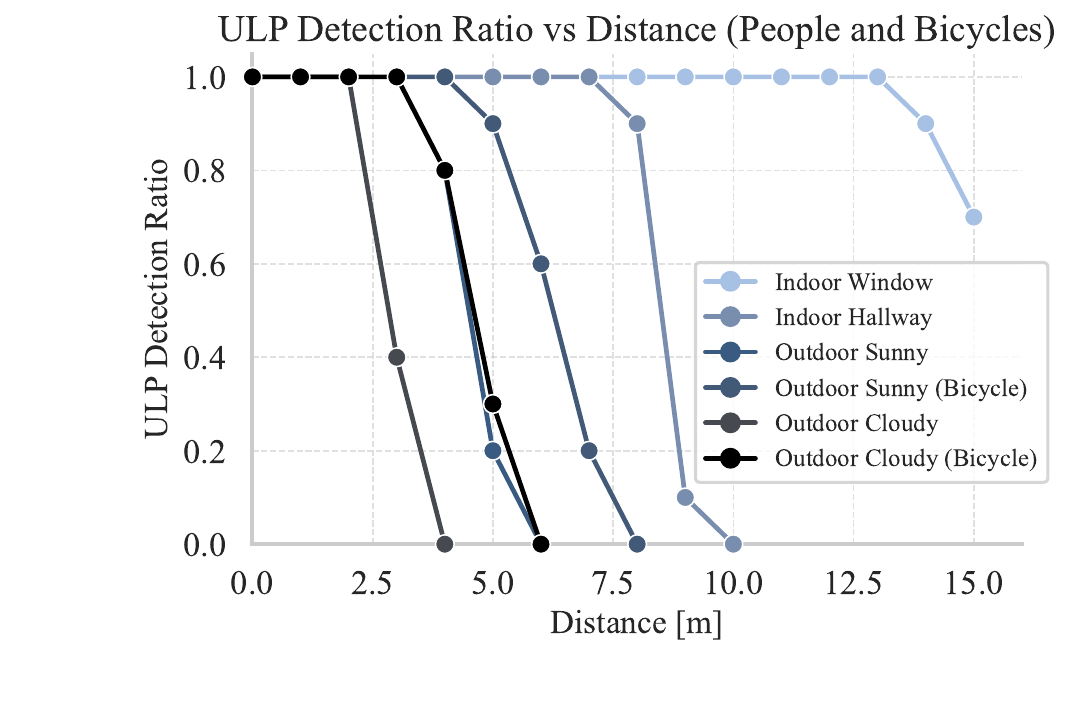}
    \caption{Evaluation of the ULP-mode capabilities of the \textsc{GENX320} event-based camera.}
    \label{fig:ulp_evaluation}
\end{figure}
\cref{fig:ulp_evaluation} visualizes the detection capability of the \textsc{GENX320} camera in \gls{ulp} mode. The indoor scenario with a window in the background yields the largest detection distance due to strong illumination changes, which are ideal for event cameras. All passes are detected up to \SI{13}{m}, while the fraction of positive detections gradually decreases to 7/10 at \SI{15}{m}. When a person\footnote{Note:  A person of height \SI{180}{cm} and shoulder-width of \SI{50}{cm} at a distance of \SI{10}{m} maps to 32 by 9 pixels, while mapping to 21 by 6 pixels at a distance of \SI{15}{m}} is passing by in an indoor hallway, the person is always detectable until \SI{7}{m}, dropping to 9/10 at \SI{8}{m} and only 1/10 at \SI{9}{m}. 
Outdoors, the detection range is significantly lower. All passes are detected up to \SI{2}{m} and \SI{3}{m} away in cloudy and sunny scenarios, respectively. By \SI{4}{m} and \SI{6}{m}, respectively, no detections occur anymore. The lower overall detection range can be explained by the relatively small change in light intensity happening outdoors. On sunny days, walking past the camera blocks direct sunlight, casting a shadow that produces a larger change in illuminance while increasing the object's apparent size. 
Similarly, for bicycles outdoors, the larger change in illuminance caused by the object's increased size and speed as it passes the camera contributes to a longer detection range.

\subsection{Multimodal Vision Node Energy Consumption}\label{sec:energy_results}
The \textsc{nRF52} is activated once the \textsc{GENX320}---operating in \gls{ulp} mode---sends an interrupt. This interrupt wakes up the \gls{mcu} , which in turn power-cycles the \textsc{GAP9} microcontroller. Once \textsc{GAP9} is initialized, the \textsc{HM0360} camera captures a raw Bayer-pattern image with a resolution of 320x240. The image is demosaiced, white-balanced, and reshaped into a quadratic image in the \textsc{hwc} format of 256x256x3, which is evaluated by the \url{TYv12\_5-1.0M} with 1 million parameters. This network was chosen over the other versions due to its comparatively good detection performance and its ability to effectively leverage available hardware acceleration.  Thereafter, \gls{nms} is applied to the output of the network. If an object of interest is detected, the bounding box information is sent back to the \textsc{nRF52} via SPI, which then forwards the extracted information via \textsc{LoRa} to a visualization frontend. Once there is no visual wake-up from the GENX320 camera, the system enters low-power mode and continues to sense the environment for visual motion. 

\begin{figure*}
    \centering
    \begin{overpic}[width=\textwidth]{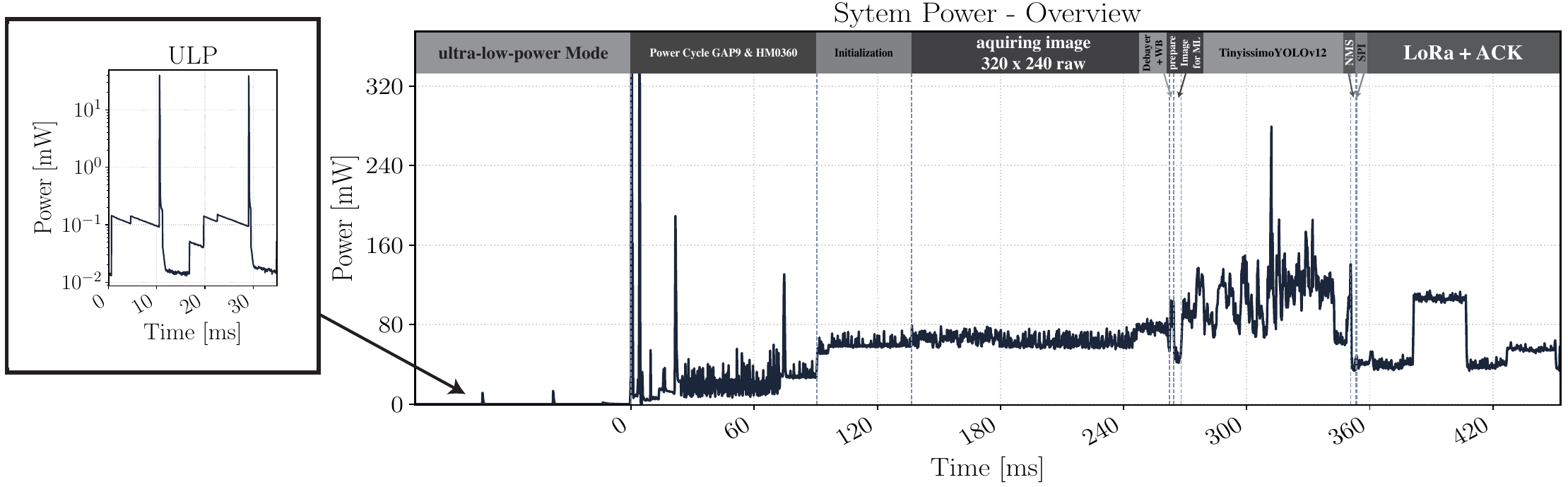}
    \put(11,5){{\scriptsize(a)}}
    \put(62.5,-1){{\footnotesize(b)}}
    \end{overpic}
    \caption{Power consumption of the system in \gls{ulp} mode (a) and in active mode (b).}
    \label{fig:power_consumption}
\end{figure*}

\cref{fig:power_consumption} visualizes the power consumption measured for the complete system. An average current of \SI{60}{\micro A} is measured at the \SI{3.7}{V} battery rail in the \gls{ulp} mode, and \SI{17.1}{mA} during the \SI{452}{ms} of active mode. The \textsc{GENX320} camera draws a static average current of \SI{22.9}{\micro A} at a voltage of \SI{2.5}{V}. The remaining power is contributed by a level-shifter that is needed for shifting the \SI{2.5}{V} signals down to the \textsc{nRF52}'s \gls{gpio} lines operating on \SI{1.8}{V}, an inverter used for booting the camera directly into the \gls{ulp} mode, a low-dropout regulator, and DC-DC inefficiencies, as well as the \textsc{nRF52} in deep sleep. The active mode includes the system and power-cycling overhead for booting the \textsc{GAP9} and \textsc{HM0360} camera, initializing the \textsc{GAP9} peripherals, and constructing the \gls{cnn}. Thereafter, the system can acquire an image. The time to first image reception is \SI{125.8}{ms} where the system consumes \SI{8.3}{mJ}. Once the image is aquired, \SI{2}{ms} are needed demosaicing the raw Bayern-pattern image and white balancing it on the cluster core of \textsc{GAP9} consuming \SI{24.5}{mA}, as well as \SI{3.7}{ms} for preparing the image to be in \gls{hwc} format of 256 x 256 x 3 requiring \SI{13.5}{mA}. The inference on the cluster core and \gls{ne16} consumes an average of \SI{28.8}{mA} over \SI{82.7}{ms}, corresponding to \SI{8.8}{mJ} of energy. A similar amount of energy is required to acquire an image. 

Lastly, \gls{nms} postprocessing requires \SI{2.5}{ms} and \SI{10.9}{mA}. Sending the result to the \textsc{nRF52} via SPI is completed within \SI{127}{\micro s} at a current of \SI{10.8}{mA}, while sending the result with LoRa consums \SI{11.1}{mA} for \SI{980}{\micro s} using a spreading factor of 7, \SI{14}{dBm} transmission power and a bandwidth of \SI{250}{kHz}. Waiting for an acknowledgement of the LoRa transmission with the same transmission specifications, the LoRa transmission completes within \SI{99.2}{ms} with an average current consumption of \SI{16.5}{mA}, requiring significantly more time. On average, the active power consumption is \SI{63.4}{mW} during \SI{452}{ms}, resulting in a total energy consumption of \SI{28.7}{mJ} for waking up, sensing the visual environment, and sending the found information to a visualization frontend. A battery with the dimensions illustrated in \cref{fig:visioniotnode_rendering} has an energy density of \SI{6.66}{kJ}. Additionally, assuming a 1\% daily activity ratio, the active mode is activated 1728 times a day. This results in an energy consumption of \SI{3.07}{J} per hour, emptying the battery within 90 days. 

\section{Conclusion}\label{sec:conclusion}
This work presents an end-to-end evaluation of an energy-efficient, always-on, multi-modal visual IoT node that demonstrates continuous environmental monitoring. The proposed design maximizes battery life while maintaining continuous visual awareness using a \textsc{GENX320} event camera in \gls{ulp} mode. Upon visual wake-up, an \acrshort{rgb} camera captures an image, which is fed into a powerful \gls{sota} TinyissimoYOLOv12 hardware-aware object detection network that analyzes the object causing the motion in front of the camera's \glspl{fov}. The extracted environmental context is then forwarded via LoRa telemetry. The system's energy consumption demonstrates the capability for a visual IoT node in the place-and-forget paradigm, for up to 3 months of battery runtime with a battery size of only \SI{1.85}{Wh}.
The \gls{ulp} mode of the \textsc{GENX320} provides continuous visual motion wake-up capability, with wake-to-motion of up to 13 meters indoors and 7 meters outdoors, while using a lens with a \gls{fov} of 104°. 

Changing the optics will result in a greater distance with a reduced \gls{fov}, thereby making the system tunable to application-specific needs.
In addition, the deployed object detection networks demonstrate attention-based object detection inference on a microcontroller-class device.  A dataset is collected to fine-tune the network for application-specific needs. The 80-class object detection network is deployed and end-to-end demonstrated, with postprocessing of the network's output using \gls{nms} completing in \SI{85.1}{ms}.

The presented architecture illustrates that a heterogeneous sensing system operating energy-proportional to activity, combined with \gls{tinyML}, improves energy efficiency without compromising context awareness.

\section*{Acknowledgments}
The authors would like to thank the students involved to this work: Linus Meier, Leo An, Niklas Weiler, Ciaran Wehrli, Jonas Gutmann, Varsha Jayaprakash, Dennis Vilgertshofer, and Tom Stein.

The authors would like to thank armasuisse Science \& Technology for funding this research.

\bibliographystyle{IEEEtranDOI}
\bibliography{IEEEabrv}

\newpage

\section*{Biography Section}
\begin{IEEEbiography}[{\includegraphics[width=1in,height=1.25in,clip,keepaspectratio]{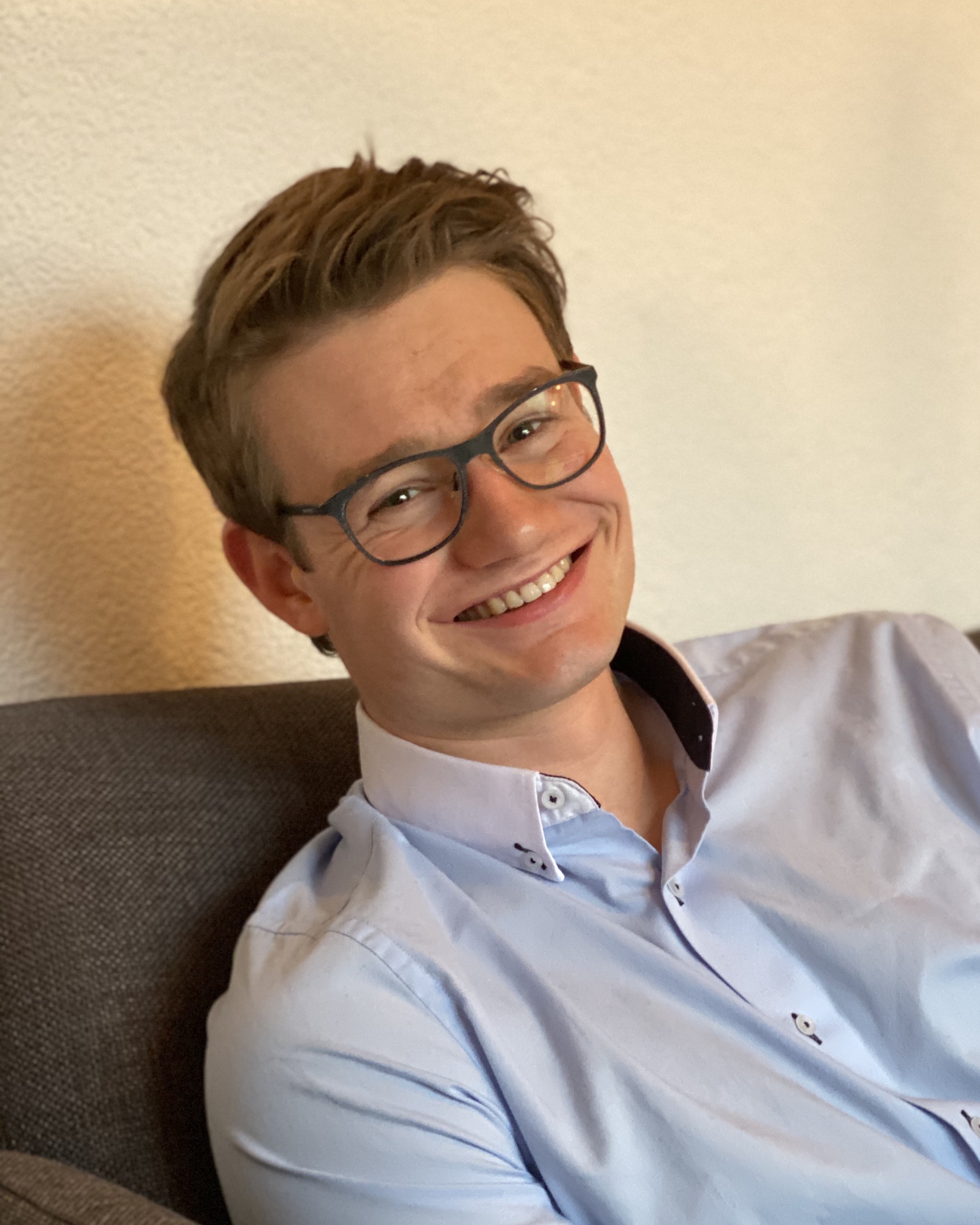}}]{Julian Moosmann}received the B.Sc. and M.Sc. degrees in electrical engineering and information technologies from ETH Zürich, Zürich, Switzerland, in 2019 and 2023, respectively. From 2022 to 2023, he was a Research Assistant at the Center for Project-Based Learning D-ITET, ETH Z\"urich, Z\"urich, Switzerland, where he is currently conducting his doctorate to pursue the degree of Doctor of Science with the Integrated Systems Laboratory in conjunction with the Center for Project-Based Learning D-ITET. 

His research interests include a combination of computer vision, event-based sensing, low-power systems, wireless sensor networks, tiny machine learning / onboard intelligence, and battery-operated distributed systems.
\end{IEEEbiography}
\vspace{-0.5cm}
\begin{IEEEbiography}[{\includegraphics[width=1in,height=1.25in,clip,keepaspectratio]{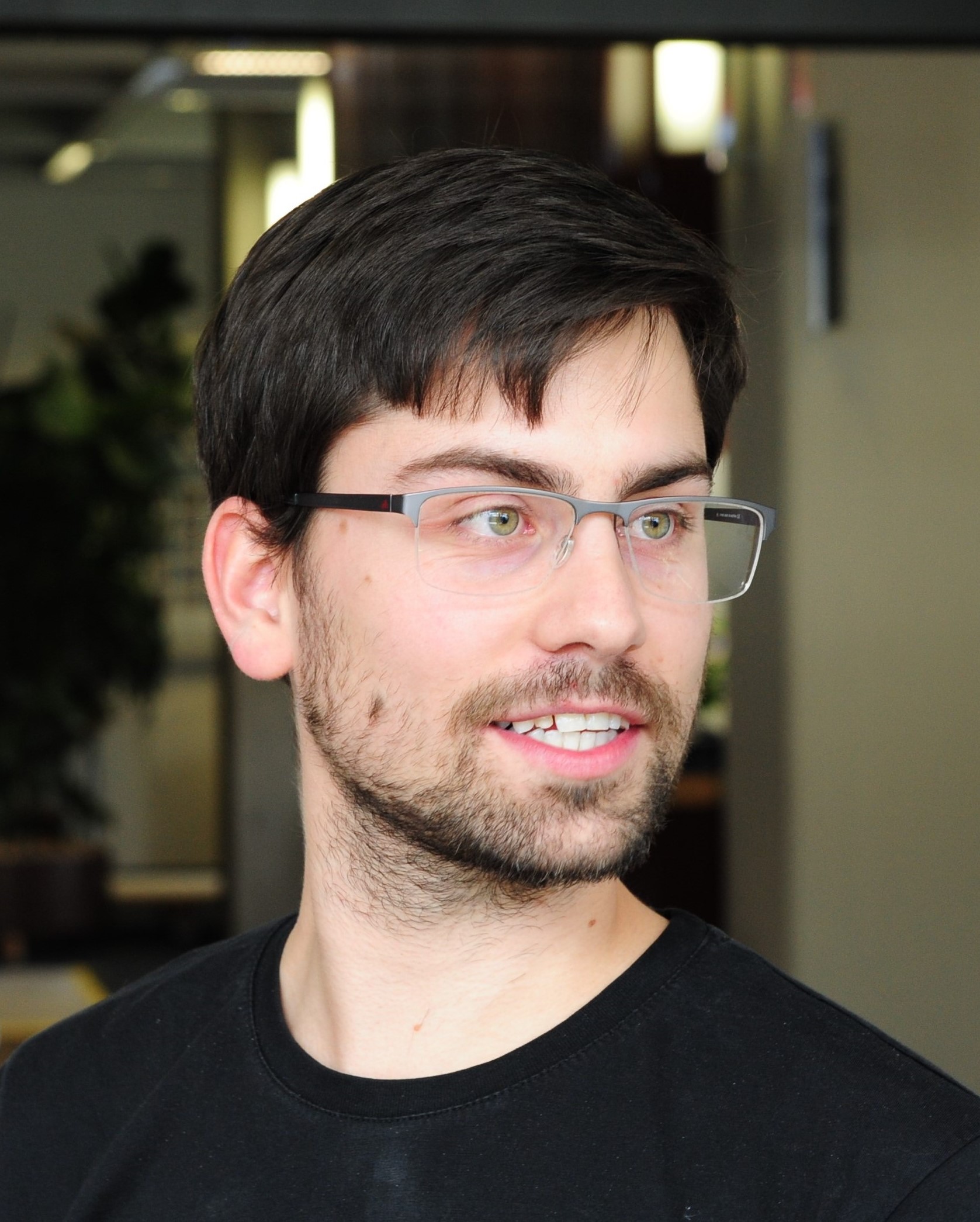}}]{Philipp Mayer} received the B.Sc. degree in electrical engineering and information technology from TU Wien, Vienna, Austria, in 2016, and the M.Sc. degree from ETH Zurich, Zurich, Switzerland, in 2018. He received the Ph.D. degree from ETH Zurich in 2022, where he was with the Integrated Systems Laboratory.\\
His research interests include low-power system design, energy harvesting, and edge computing. \\
Dr. Mayer was a recipient of the Best Paper Award at the 2017 IEEE International Workshop on Advances in Sensors and Interfaces and the Best Student Paper Award at the 2018 IEEE Sensors Applications Symposium. Beyond his particular area of expertise, he was granted the Best Poster Award in the 2018 IOP Workshop in Devices, Materials and Structures for Energy Harvesting and Storage. In 2019, he founded Mayer Engineering and Consulting, intending to connect academics with industrial expertise.
\end{IEEEbiography}
\vspace{-0.5cm}
\begin{IEEEbiography}[{\includegraphics[width=1in,height=1.25in,clip,keepaspectratio]{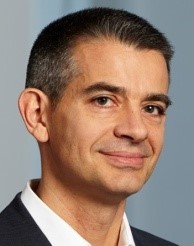}}]{Luca Benini}
(Fellow, IEEE) received the Ph.D. degree in electrical engineering from Stanford University, Stanford, CA, USA, in 1997.\\
He has served as the Chief Architect of the Platform2012/STHORM Project with STMicroelectronics, Grenoble, France, from 2009 to 2013. Currently, he holds the Chair of Digital Circuits and Systems at ETH Zurich, Zurich, Switzerland, and is a Full Professor at the University of Bologna, Bologna, Italy. He has published more than 1000 peer-reviewed articles and five books. His current research interest includes energy-efficient computing systems' design from embedded to high performance.\\
Dr. Benini is a fellow of the ACM and a member of the Academia Europaea. He was a recipient of the 2016 IEEE CAS Mac Van Valkenburg Award and the 2019 IEEE TCAD Donald O. Pederson Best Article Award.
\end{IEEEbiography}
\vspace{-0.5cm}
\begin{IEEEbiography}[{\includegraphics[width=1in,height=1.25in,clip,keepaspectratio]{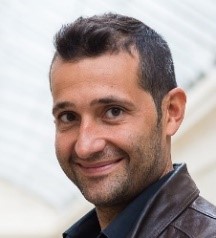}}]{Michele Magno}
(Fellow, IEEE) received he master's and Ph.D. degrees in electronic engineering from the University of Bologna, Bologna, Italy, in 2004 and 2010, respectively. \\
Currently, he is a Senior Researcher at ETH Zurich, Zurich, Switzerland, where he is the Head of the Project-Based Learning Center. He has collaborated with several universities and research centers, such as Mid University Sweden, where he is a Guest Full Professor. He has published more than 150 articles in international journals and conferences, in which he got multiple best paper and best poster awards. The key topics of his research are wireless sensor networks, wearable devices, machine learning at the edge, energy harvesting, power management techniques, and extended lifetime of battery-operated devices.
\end{IEEEbiography}

\vfill

\end{document}